\documentclass[a4paper,11pt]{article}
\pdfoutput=1
\usepackage{jcappub}
\usepackage{mathtools}
\graphicspath{{plots/}}

\newcommand{\gs}{g_\text{eff}}
\newcommand{\gss}{h_\text{eff}}
\newcommand{\Trh}{T_\text{rh}}

\newcommand{\Tmax}{T_\text{max}}
\newcommand{\sv}{\langle\sigma v\rangle}
\newcommand{\rR}{\rho_R}
\newcommand{\rp}{\rho_\phi}

\title{Kination and the Inert Doublet Model}
\author[a]{Geneviève Bélanger,}
\author[b]{Nicolás Bernal,}
\author[c]{\\Andreas Goudelis,}
\author[d]{Alexander Pukhov}

\affiliation[a]{LAPTh, CNRS, Université Savoie Mont-Blanc\\
9 Chemin de Bellevue, 74940 Annecy, France}
\affiliation[b]{New York University Abu Dhabi\\
PO Box 129188, Saadiyat Island, Abu Dhabi, United Arab Emirates}
\affiliation[c]{Laboratoire de Physique de Clermont Auvergne (UMR 6533), CNRS/IN2P3\\
Univ. Clermont, Auvergne, 4 Av. Blaise Pascal, F-63178 Aubière Cedex, France}
\affiliation[d]{Skobeltsyn Institute of Nuclear Physics, Moscow State University\\
Moscow 119992, Russia}

\emailAdd{belanger@lapth.cnrs.fr}
\emailAdd{nicolas.bernal@nyu.edu}
\emailAdd{andreas.goudelis@clermont.in2p3.fr}
\emailAdd{alexander.pukhov@gmail.com}

\abstract{The inert doublet model is a two-Higgs-doublet extension of the standard model that provides a minimal and versatile framework for frozen-out dark matter. Assuming standard cosmology, if the dark matter mass ranges between approximately 120 GeV and 500 GeV then it turns out to be underabundant, as gauge interactions render its annihilation too efficient. In this work, we show that this mass window becomes allowed in cosmological scenarios where dark matter freeze-out occurs during a period with a stiff equation of state, $w > 1/3$, such as kination. This predictive setup satisfies all current experimental constraints while remaining within the reach of upcoming detection efforts.}

\begin{document}
\begin{flushright}
\end{flushright}
\maketitle

\section{Introduction}
Despite significant progress over the past several decades, dark matter (DM) remains one of the most compelling open problems in particle physics, cosmology, and astrophysics~\cite{Planck:2018vyg}. From a particle physics perspective, the Inert Doublet Model (IDM) stands out as one of the simplest and most economical extensions of the standard model (SM) that tries to address the DM question~\cite{Deshpande:1977rw, Ma:2006km, Barbieri:2006dq}.

Due to its rich phenomenology, the IDM has attracted considerable attention in the literature. Its DM candidate encompasses all the fundamental mechanisms by which the observed relic abundance can be produced in weakly interacting massive particle (WIMP)~\cite{Arcadi:2024ukq} scenarios. In particular, the entire observed relic density can be obtained by appropriately tuning the model couplings, by exploiting resonant annihilation effects or moving away from them, and by invoking co-annihilation processes with other inert states, within a wide range of masses. However, for DM masses between $\sim 120$~GeV and $\sim 500$~GeV the model has since long been considered incapable of explaining the observed DM abundance in the Universe, as the predicted relic density is too low due to very efficient DM annihilations mediated by (fixed!) SM gauge couplings.

In this work, we show that this conclusion, which relies on the assumption of the standard cosmological scenario, does not hold in more general settings where the DM relic abundance is produced during non-standard cosmological epochs. It is worth emphasizing that the standard cosmological paradigm -- namely that before the onset of the Big Bang nucleosynthesis era $i)$ the SM entropy is conserved, $ii)$ the Hubble expansion rate is driven by the SM energy density, and $iii)$ cosmic reheating occurs instantaneously at a sufficiently high temperature -- is not guaranteed. Indeed, departures from this framework are generic in cosmology and are well motivated in ultraviolet-complete scenarios~\cite{Allahverdi:2020bys, Batell:2024dsi}.

As a representative example, we consider the case in which DM freeze-out occurs during a stiff era, where the expansion rate of the Universe is dominated by a component with an equation-of-state parameter $w = 1$, as in kination~\cite{Zeldovich:1961sbr, Spokoiny:1993kt, Joyce:1996cp, Ferreira:1997hj}. This scenario could arise, for example, in the context of axion kinetic misalignment~\cite{Chang:2019tvx, Co:2019jts, Barman:2021rdr}, or in quintessence inflation with a low-reheating temperature~\cite{Dolgov:1989us, Traschen:1990sw, Spokoiny:1993kt, Kofman:1994rk, Kofman:1997yn, Peebles:1998qn, Bernal:2020bfj, Barman:2025lvk}. Our conclusions, however, extend to more general scenarios with $w>1/3$, in which the component driving the expansion of the Universe does not need to decay or annihilate into SM states.

The paper is organized as follows: in Section~\ref{sec:idm} we review the IDM and summarize the main experimental constraints it is subject to, with particular emphasis on those arising from direct and indirect DM detection. In Section~\ref{sec:DM} we discuss DM production, first within the standard cosmological framework and then in the presence of kination and more general non-standard cosmological eras. Finally, our conclusions are presented in Section~\ref{sec:conclu}.

\section{The Inert Doublet Model} \label{sec:idm}
The IDM is a simple extension of the SM by an exotic complex scalar $SU(2)_L$ doublet $\Phi$. A discrete $\mathbb{Z}_2$ symmetry is imposed at the Lagrangian level, under which $\Phi$ is taken to be odd, whereas the SM particles are even. Then, as long as this symmetry remains unbroken, the exotic scalar doublet is inert in the sense that its component fields only couple in pairs with the SM particles, and the lightest component is stable, rendering it a viable DM candidate.

Under these assumptions, the only part of the SM Lagrangian that is modified at tree-level is the scalar potential, which in the IDM reads
\begin{equation}
    V_0 ~=~ \mu_1^2 |H|^2  + \mu_2^2|\Phi|^2 + \lambda_1 |H|^4+ \lambda_2 |\Phi|^4 + \lambda_3 |H|^2| \Phi|^2 + \lambda_4 |H^\dagger\Phi|^2 + \frac{\lambda_5}{2} \Bigl[ (H^\dagger\Phi)^2 + \mathrm{H.c.} \Bigr],
    \label{Eq:TreePotential}
\end{equation}
where $H$ is the usual SM Higgs doublet. In everything that follows, the couplings $\lambda_i$ will be taken to be real. Upon electroweak symmetry breaking, the two doublets can be expanded as
\begin{equation}
	H ~=~ \left( \begin{array}{c} G^+ \\ \frac{1}{\sqrt{2}}\left(v+h+\mathrm{i}G^0\right) \end{array} \right),
	\qquad
	\Phi ~=~ \left( \begin{array}{c} H^+\\ \frac{1}{\sqrt{2}}\left(H^0+\mathrm{i}A^0\right) \end{array} \right),
\end{equation}
where $v \simeq 246$~GeV is the usual vacuum expectation value of the neutral component of $H$, $h$ is the physical SM Higgs-boson and $G^0/G^{\pm}$ are Goldstone bosons. The inert sector, in turn, consists of a neutral CP-even scalar $H^0$, a pseudo-scalar $A^0$, and a pair of charged scalars $H^\pm$. In total, the IDM introduces five new free parameters
\begin{equation}
    \left\{\mu_2, ~~\lambda_2, ~~ \lambda_3, ~~ \lambda_4, ~~ \lambda_5 \right\}.
    \label{eq:lambdas}
\end{equation}
The tree-level scalar masses can be read off the potential of Eq.~\eqref{Eq:TreePotential} and are given by
\begin{align}
	m_{H^0}^2 &= \mu_2^2 + \lambda_L\, v^2, \label{Eq:mH0tree} \\
	m_{A^0}^2 &= \mu_2^2 + \lambda_S\, v^2, \\
	m_{H^\pm}^2 &= \mu_2^2 + \frac12\, \lambda_3\, v^2,
\end{align}
where we have defined
\begin{align}
	\lambda_L &\equiv \frac12 \left( \lambda_3 + \lambda_4 + \lambda_5 \right), \\
	\lambda_S &\equiv \frac12 \left( \lambda_3 + \lambda_4 - \lambda_5 \right).
\end{align}
These combinations correspond, respectively, to the coupling of a pair of $H^0$ or $A^0$ particles to the SM Higgs boson. These relations can be used to trade the parameters in Eq.~\eqref{eq:lambdas} for the physically more intuitive ones
\begin{equation}
 	\left\{ m_{H^0}, ~~ m_{A^0}, ~~ m_{H^\pm}, ~~ \lambda_L, ~~ \lambda_2 \right\}.
	\label{eq:masses}
\end{equation} 
The IDM was first proposed as a DM model in Refs.~\cite{Ma:2006km, Barbieri:2006dq}, almost thirty years after its original introduction in the context of electroweak symmetry breaking studies~\cite{Deshpande:1977rw}. Its relic abundance was explored in detail in Refs.~\cite{LopezHonorez:2006gr, LopezHonorez:2010tb} and, eventually, in Ref.~\cite{Goudelis:2013uca}, following the LHC discovery of the Higgs boson. It contains two viable candidates for DM, $H^0$ and $A^0$, with essentially identical phenomenologies. For concreteness, in everything that follows, we will focus on the case of $H^0$. 

Given the fact that $\Phi$ is charged under the electroweak gauge group, DM particles thermalize with the SM in the early Universe, which implies that DM undergoes thermal freeze-out. Concretely, DM particles can interact with the SM in the following set of ways:
\begin{itemize}
    \item They can (co-)annihilate into pairs of SM fermions through the SM (gauge) Higgs bosons. These processes are mostly important in the low-mass regime of the model, as will be described below.
    \item They can annihilate into pairs of Higgs bosons, through interactions involving the scalar potential parameters $\lambda_L$ and $\lambda_1$.
    \item They can (co-)annihilate into pairs of SM gauge bosons. These interactions involve either pure gauge couplings $g$ or the product $g\times\lambda_L$.
\end{itemize}
As described in detail in Refs.~\cite{LopezHonorez:2006gr, LopezHonorez:2010tb, Goudelis:2013uca, Eiteneuer:2017hoh}, the IDM parameter space can be divided into four broad regions, depending on the DM mass: 
\begin{itemize}
    \item For $m_{H^0} \lesssim 65$~GeV, the DM relic density is in principle driven by (co-)annihilations into SM fermions. Given LEP2 bounds on the mass of the next-to-lightest neutral $\mathbb{Z}_2$-odd state~\cite{Pierce:2007ut, Lundstrom:2008ai}, in practice coannihilations turn out to be irrelevant in this mass regime. Annihilation processes, in turn, are controlled by $\lambda_L$, the coupling of a pair of $H^0$ to the SM Higgs boson, which also determines the WIMP-nucleon scattering cross-section at tree-level. With the exception of the Higgs resonance $m_{H^0} \simeq m_h/2 \simeq 62.5$~GeV, this region is now essentially excluded by direct detection experiments~\cite{Eiteneuer:2017hoh}.
    \item For $m_{H^0} \gtrsim 65$~GeV annihilations into (potentially virtual) gauge boson pairs start becoming relevant and, eventually, completely dominate as the $W^+ W^-$ threshold is approached from below. In fact, this annihilation channel becomes so efficient that $\lambda_L$ is restricted to increasingly negative values in order to achieve a cancelation between the four-vertex and $\mathbb{Z}_2$-odd state-mediated Feynman diagrams (which only depend on the weak gauge coupling $g$) on one hand and, on the other hand, the Higgs-mediated one (which, in turn, depends on the product $g\times\lambda_L$, for a detailed discussion see Ref.~\cite{LopezHonorez:2010tb}). 
    \item As the $hh$ threshold is approached from below, and for such large negative values of $\lambda_L$, annihilation into Higgs boson pairs becomes efficient. Hence, for $m_{H^0} \gtrsim 120$~GeV it turns out to be impossible to reproduce the observed DM abundance in the Universe~\cite{Goudelis:2013uca}: the predicted relic density turns out to be too low either due to strong annihilation into gauge bosons (small or positive $\lambda_L$ values) or into Higgs boson pairs (large negative $\lambda_L$ values). Note that for a non-vanishing $\lambda_L$ value, annihilation into $t\bar{t}$ pairs can also become relevant~\cite{LopezHonorez:2010tb}.
    \item Lastly, once $m_{H^0} \gtrsim 500$~GeV is reached, destructive interference between Feynman diagrams can become efficient enough so that the  $\Omega h^2$ constraint can be satisfied again. In this regime, and for a given DM mass, the predicted relic density depends on the one hand on the mass difference between $H^0$ and the heavier states (with the annihilation cross-section into gauge boson final states increasing for larger values of $\Delta m$) and, on the other hand, on the value of $\lambda_L$ (with negative values of $\lambda_L$ being partly capable of counter-balancing a larger mass splitting but, at the same time, giving rise to annihilations into Higgs boson pairs).
\end{itemize}
In summary, for $m_{H^0} \gtrsim 500$~GeV and for a given value of $\lambda_L$ it is always possible to choose the mass difference between $H^0$ and $A^0/H^\pm$ such that the IDM fits the entire cosmic DM abundance. For $120 \ {\rm GeV } \lesssim m_{H^0} \lesssim 500$~GeV, on the other hand, the predicted DM relic density turns out to be too low. It is this intermediate regime that we will mostly focus on in everything that follows.

\begin{figure}[t!]
    \def\sepf{0.496}
    \centering
    \includegraphics[width=\sepf\columnwidth]{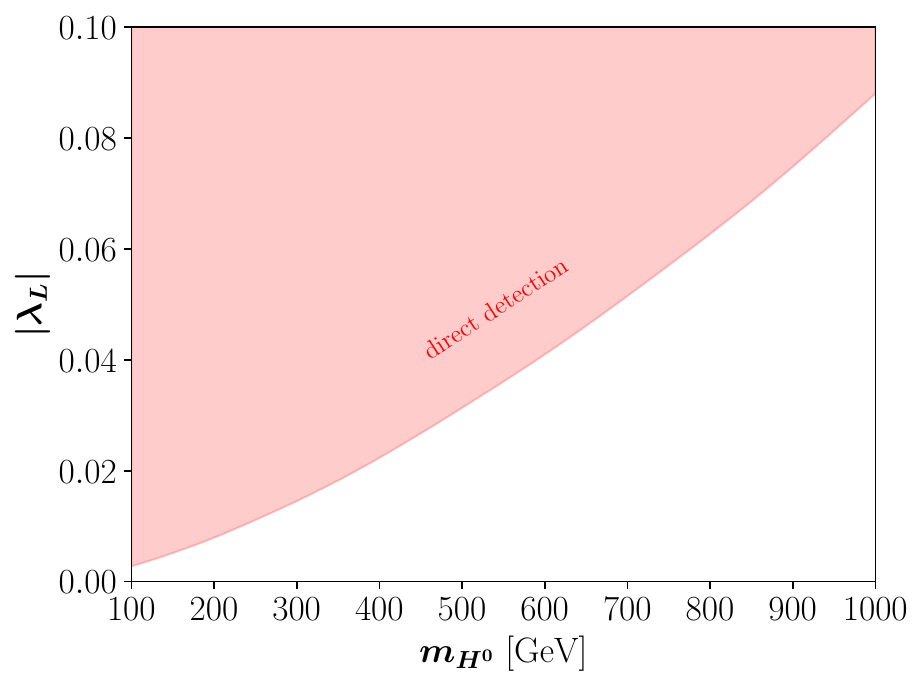}
    \includegraphics[width=\sepf\columnwidth]{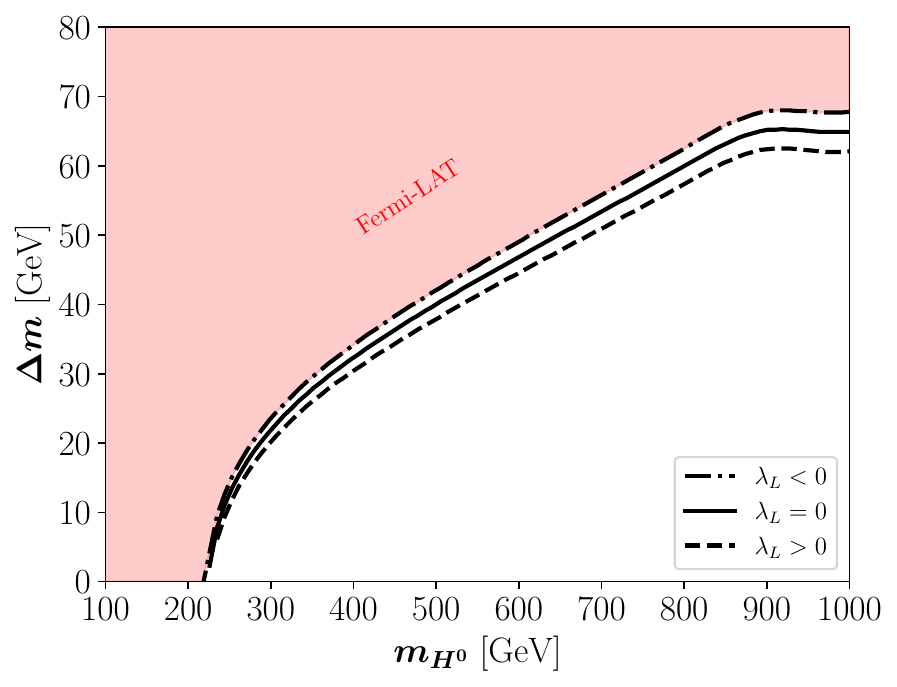}
    \caption{Left: {\it Maximum} value of $|\lambda_L|$ allowed by DM direct detection. Right: {\it Maximum} mass splitting $\Delta m \equiv m_{A^0} - m_{H^0}$ (with $m_{A^0} = m_{H^\pm}$) allowed by DM indirect detection, for the maximal and minimal value of $\lambda_L$ (cf. left panel), and $\lambda_L = 0$.}
    \label{fig:detection}
\end{figure}

In practice, we will study three sets of scenarios: the first is the case $\lambda_L = 0$. This scenario is of particular interest, since the WIMP-nucleon scattering cross-section vanishes at tree-level.\footnote{Note, however, that one-loop electroweak corrections may bring this scenario within the reach of currently running direct detection experiments in the near future; see, for example, Ref.~\cite{Klasen:2013btp}.} For the second and third ones, we will choose the absolute value of $\lambda_L$ such that direct detection constraints (which, at tree-level, do not depend on the sign of $\lambda_L$) are satisfied assuming that the IDM accounts for the entire observed DM abundance in the Universe. The corresponding constraint is shown in the LHS panel of Fig.~\ref{fig:detection}, in which we compute the {\it maximum} value of $|\lambda_L|$ allowed by the latest limits on direct detection of DM from LZ~\cite{LZ:2024zvo}. As we can see, to avoid these bounds, a fairly suppressed coupling $|\lambda_L|$ is required, which lies in the range of $10^{-2}$ for $100 \ {\rm GeV } \lesssim m_{H^0} \lesssim 500$~GeV. 

For each of the three cases ($\lambda_L$ positive, negative, or zero), we will further distinguish two benchmarks for the mass difference $\Delta m \equiv m_{A^0} - m_{H^0}$ between $m_{H^0}$ and $m_{A^0} = m_{H^\pm}$: first, a minimal mass difference of $\Delta m_\text{min} = 1$~GeV. This choice ensures that the heavier charged inert state decays promptly and easily escapes  constraints from LLP searches at the LHC, namely searches for disappearing tracks and heavy stable charged particles which  exclude $\Delta m \approx 130-300~{\rm MeV}$ for $100 \ {\rm GeV } \lesssim m_{H^0} \lesssim 500$~GeV~\cite{Heisig:2018kfq, Alguero:2021dig}. Second, the maximal mass difference that is allowed from indirect searches for gamma-rays in dwarf spheroidal galaxies with Fermi. The corresponding results are shown in the right panel of Fig.~\ref{fig:detection}, in which we compute the {\it maximum} value $\Delta m_\text{max}$ of the mass splitting $\Delta m$ allowed by Fermi-LAT, fixing $\lambda_L$ to its maximum value, its minimum value, or zero, as shown in the left panel of Fig.~\ref{fig:detection}. The Fermi-LAT limits are obtained by an implementation of the method described in Ref.~\cite{Calore_2021_5592836} included in {\tt micrOMEGAs}~\cite{Alguero:2023zol}. Assuming an NFW profile~\cite{Navarro:1996gj}, this code uses a data-driven approach to estimate the background and derive conservative but robust limits on the thermally averaged DM pair-annihilation cross-section based on 10 years of Fermi-LAT data from observations of 25 Milky Way's dwarf spheroidal galaxies~\cite{Fermi-LAT:2013sme, Calore:2018sdx, Alvarez:2020cmw}. It is interesting to note that small mass splittings (at the few percent level) are allowed, with only a mild dependence on the value and the sign of $\lambda_L$ -- the latter being severely constrained to small values by direct detection, as discussed previously. Moreover, it should be noted that the data exclude DM masses smaller than $m_{H^0} \lesssim 225$~GeV.

Lastly, additional constraints on the IDM parameter space arise due to theoretical consistency requirements (stability of the electroweak vacuum, perturbativity, and perturbative unitarity) as well as from potential contributions to the $S$, $T$ and $U$ oblique parameters. We have computed these constraints with the help of the {\tt 2HDMC} package~\cite{Eriksson:2009ws} and have found them to be easily satisfied throughout the parameter space that we will consider, with one notable exception for negative $\lambda_L$ values. We will comment on this case in Appendix~\ref{app:more}.

\section{The dark matter relic density} \label{sec:DM}
Let us now turn to the DM relic density predicted in the IDM. The evolution of the DM number density $n$ can be computed using the Boltzmann equation
\begin{equation} \label{eq:BEDM}
    \frac{dn}{dt} + 3\, H\, n = -\sv \left(n^2 - n_\text{eq}^2\right),
\end{equation}
where $n_\text{eq}$ corresponds to the equilibrium DM number density, $\sv$ to its total annihilation cross section into a pair of SM states and $H$ to the Hubble expansion rate. Once a specific model is chosen -- the IDM in this case -- all the particle physics is fixed. 

However, it is important to emphasize that in order to solve Eq.~\eqref{eq:BEDM} it is also necessary to make {\it assumptions} about the evolution of the cosmological background. In fact, there are large uncertainties on the cosmological history of the Universe between the end of cosmic inflation and the onset of BBN~\cite{Allahverdi:2020bys, Batell:2024dsi}. The most commonly adopted approach is to assume that in this era the energy density of the Universe was dominated by SM radiation; however, this may have not been the case. In what follows, we break this assumption down precisely. Concretely, the WIMP DM production will be studied \textit{i}) in a conventional SM radiation dominated background, and \textit{ii}) in kination and kination-like scenarios. In practice, all numerical results have been computed with new routines that have been implemented in {\tt micrOMEGAs} as described in the Appendix~\ref{app:micromegas}.

\subsection{Radiation domination}
If, following post-inflationary reheating, the Universe is dominated by SM radiation, the Hubble expansion rate $H$ is given by
\begin{equation}
    H^2 = \frac{\rho_R}{3\, M_P^2}\,,
\end{equation}
where the SM energy density reads
\begin{equation}
    \rho_R(T) = \frac{\pi^2}{30}\, \gs(T)\, T^4\,,
\end{equation}
with $\gs(T)$ being the effective number of degrees of freedom contributing to the SM energy density, $T$ the temperature of the SM photons, and $M_P \simeq 2.4 \times 10^{18}$~GeV the reduced Planck mass.

\begin{figure}[t!]
    \def\sepf{0.496}
    \centering
    \includegraphics[width=\sepf\columnwidth]{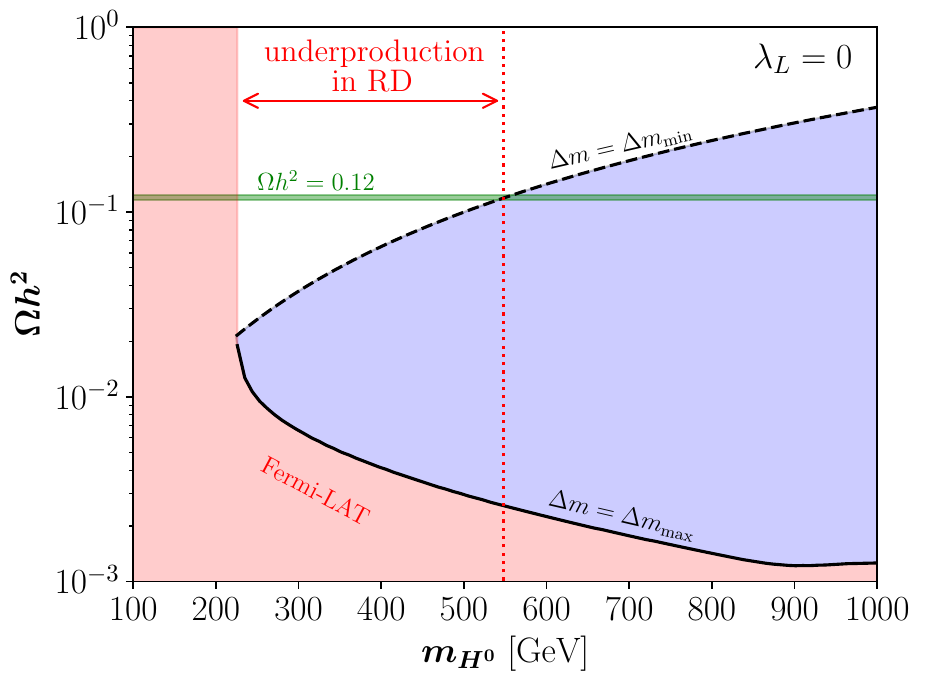}
    \caption{DM relic abundance for $\lambda_L = 0$ and different values of the mass splitting $\Delta m$. The two black lines bracket the region in blue that can be achieved in the standard cosmological scenario. The red region is in conflict with Fermi-LAT data.}
    \label{fig:RD}
\end{figure}

In Fig.~\ref{fig:RD} we show the DM relic abundance that can be generated for $\lambda_L = 0$ as a function of the DM mass. The upper dashed and lower solid black lines correspond, respectively, to the minimum ($1$ GeV) and the maximum (direct/indirect detection-inferred) mass splittings that we consider as described in Section~\ref{sec:idm}. Therefore, the blue area between these curves brackets the available parameter space. 

As can be deduced from Fig.~\ref{fig:detection}, indirect detection excludes DM masses smaller than $m_{H^0} \lesssim 225$~GeV. Moreover, in Fig.~\ref{fig:RD} we observe that the model typically predicts an under-abundance of DM $\Omega h^2 \ll 0.12$, in line with our previous discussion. In particular, for 225~GeV~$\lesssim m_{H^0} \lesssim 550$~GeV DM is {\it always} underproduced, regardless of the mass splitting $\Delta m$. That implies that, within the assumption of a SM-dominated cosmological background, the IDM cannot explain the entire observed DM cosmic abundance. Higher masses corresponding to $m_{H^0} \gtrsim 550$~GeV are viable, as long as the mass splitting is small enough. Note also that finite values for $\lambda_L$ yield fairly similar results, which we present for completeness in Appendix~\ref{app:more} (cf. upper panels of Fig.~\ref{fig:kination2}).

\subsection{Kination and beyond}
Alternatively, the total energy density of the Universe could, at some point during the cosmic evolution, have been dominated by an extra component $\phi$. The corresponding Hubble expansion rate now contains two contributions
\begin{equation}
    H^2 = \frac{\rho_R + \rho_\phi}{3\, M_P^2}\,.
\end{equation}
The second term, which in this paper we assume to come from the contribution of a fluid $\phi$ with an equation-of-state parameter $w > 1/3$, can be conveniently parameterized as
\begin{equation}
    \rho_\phi(T) = \rho_R(\Trh) \left(\frac{T}{\Trh}\right)^{3(1+w)},
    \label{eq:rhophi}
\end{equation}
with the reheating temperature $T = \Trh$ being the temperature at which the two contributions are equal. The case $w = 1$ corresponds to the so-called ``kination'' scenario~\cite{Zeldovich:1961sbr, Spokoiny:1993kt, Joyce:1996cp, Ferreira:1997hj}, however, higher values $w > 1$ can also occur~\cite{Choi:1999xn, Khoury:2001wf, Gardner:2004in, DEramo:2017gpl}.

At temperatures much higher than $\Trh$, the Hubble rate is dominated by $\phi$ and the Universe expands faster than in the case of SM radiation domination, while for lower temperatures the standard cosmological scenario is recovered:
\begin{equation}
    H(T) \simeq \frac{\pi}{3}\, \sqrt{\frac{\gs}{10}}\, \frac{T^2}{M_P} \times
    \begin{dcases}
        \left(\frac{T}{\Trh}\right)^\frac{3 w - 1}{2} & \text{ for } T \geq \Trh\,,\\
        \ \ \ \ 1 & \text{ for } \Trh \geq T\,.
    \end{dcases}
\end{equation}
We note that as $w > 1/3$, $\phi$ is diluted faster than free radiation and therefore does not have to decay into SM particles to give rise to a SM-dominated Universe. The domination of SM radiation in the total energy budget is required at low temperatures in order to reproduce a successful Big Bang nucleosynthesis era, which imposes a bound $\Trh \gtrsim 4$~MeV~\cite{Kawasaki:2000en, Hannestad:2004px, Cyburt:2015mya, deSalas:2015glj}. In contrast, the maximum temperature $\Tmax$ reached by the SM thermal bath during the $\phi$-dominated era is bounded from above by the maximum inflationary scale $H_I \simeq 2.0 \times 10^{-5}~M_P$~\cite{BICEP:2021xfz}, see Appendix~\ref{app:Tmax}. In the following, we will assume a value for $\Tmax$ higher than the temperature at which the DM freeze-out occurs, so that the DM production becomes {\it independent} of $\Tmax$.

Qualitatively, as long as $\Trh$ is not much larger than the DM standard radiation domination freeze-out temperature, we expect the existence of such a transient epoch of faster cosmic expansion to modify the evolution of the DM yield of the IDM (or any frozen-out DM candidate, for that matter, see also Ref.~\cite{DEramo:2017gpl}). Concretely, a faster expansion rate is expected to lead to an earlier freeze-out, which, in turn, means that the predicted abundance of DM is expected to increase.

\begin{figure}[t!]
    \def\sepf{0.496}
    \centering
    \includegraphics[width=\sepf\columnwidth]{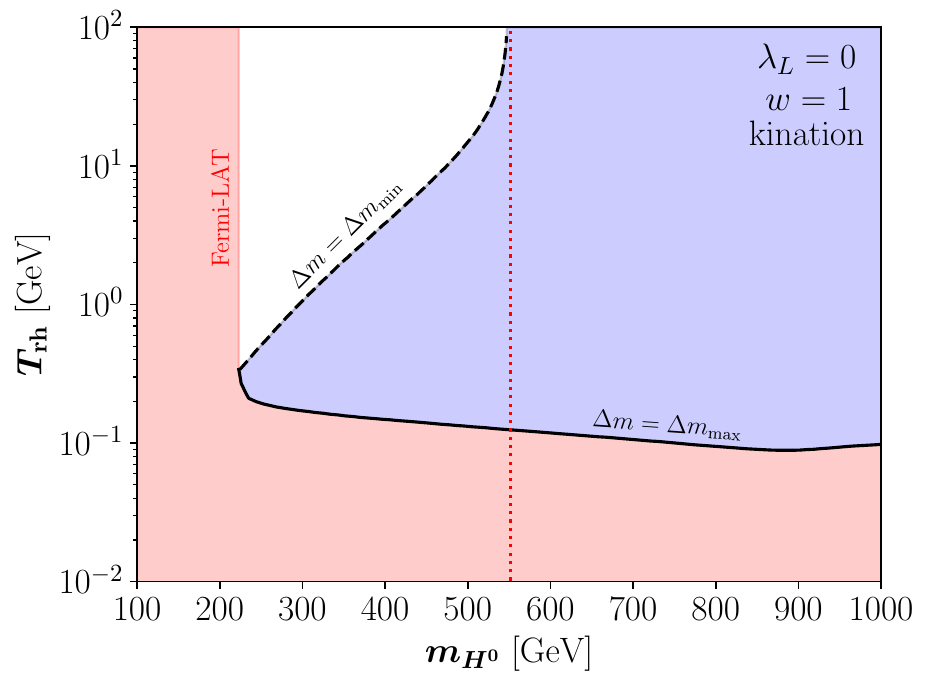}
    \includegraphics[width=\sepf\columnwidth]{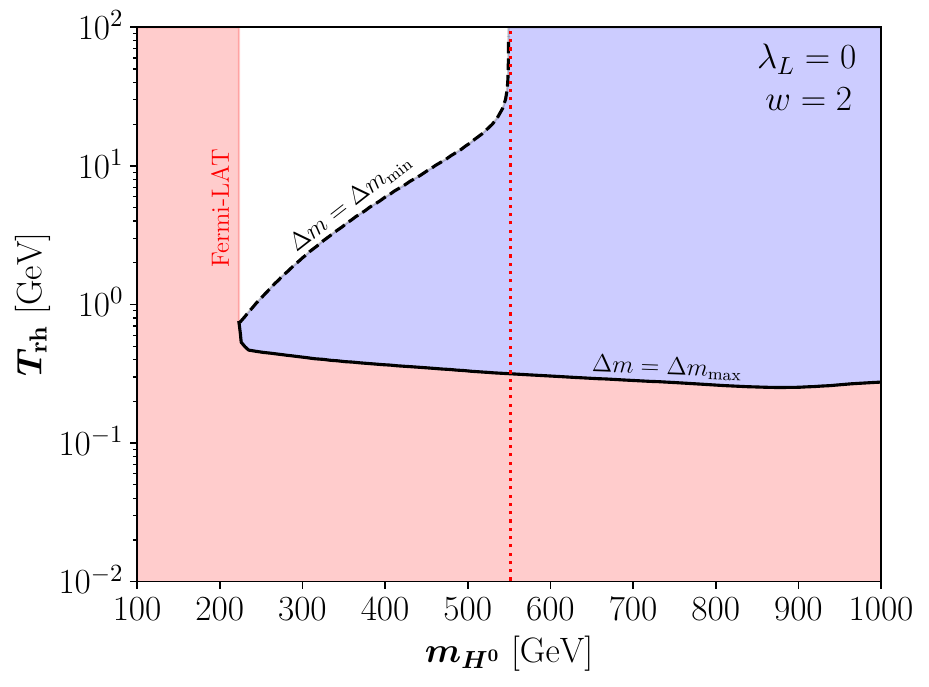}
    \caption{Values of $\Trh$ required to explain the entire observed DM abundance in the kination scenario, for $\lambda_L = 0$ and different values of the mass splitting $\Delta m$. The two black lines bracket the region in blue corresponding to the different mass splittings. The red region is in conflict with Fermi-LAT data. The left panel corresponds to kination ($w=1$) while the right panel to $w=2$.}
    \label{fig:kination}
\end{figure} 
Indeed, in Fig.~\ref{fig:kination} we show the reheating temperature $\Trh$ required to fit the entire observed DM abundance as a function of the DM mass, for the benchmark $\lambda_L = 0$, in the case of $w = 1$ (kination - left panel) and $w = 2$ (right panel). The black solid and dashed lines correspond, respectively, to the maximum and minimum values of the mass splitting $\Delta m$: the blue area contained by these two lines brackets all possible mass splittings. Several comments are in order: $i)$ Small DM masses $m_{H^0} \lesssim 225$~GeV are excluded by indirect detection. $ii)$ The mass window 225~GeV~$\lesssim m_{H^0} \lesssim 550$~GeV opens up for a reheating temperature between $\sim 100$~MeV and $\sim 100$~GeV. This effect constitutes the main result of our study. $iii)$ Higher masses $m_{H^0} \gtrsim 550$~GeV are compatible with a low reheating temperature. However, this is \textit{not} required if the mass splitting is small enough; after all, within this mass range the IDM can account for the entire DM abundance in the Universe even within standard radiation domination. $iv)$ Very small values $\Trh \lesssim 100$~MeV are not viable as they would require a mass splitting larger than the upper limit imposed by indirect detection. Lastly, finite values of $\lambda_L$ give very similar results which can be found in the Appendix~\ref{app:more} (cf. lower panels of Fig.~\ref{fig:kination2}). Finally, $v)$ higher values of $w$ correspond to a faster expansion rate, and therefore shorter non-standard eras are required leading to higher values for the reheating temperature; cf. the right panel of Fig.~\ref{fig:kination}.

\section{Conclusions} \label{sec:conclu}
In this work, we revisited the Inert Doublet Model (IDM) as a framework for WIMP dark matter (DM), emphasizing the crucial role played by the underlying cosmological history in determining the viability of its parameter space. Although the IDM is often regarded as being excluded in the intermediate DM mass range between approximately 120–500~GeV due to overly efficient DM annihilations mediated by standard model gauge interactions, we have shown that this conclusion is not generic, but only holds within the standard cosmological scenario of uninterrupted SM radiation domination.

By considering non-standard cosmological histories in which the expansion rate of the Universe is enhanced, and in particular a stiff (kination-like) era characterized by an equation-of-state parameter $w > 1/3$, we have demonstrated that the DM relic abundance can be significantly modified. In such scenarios, thermal freeze-out occurs earlier, leading to an increased relic density that can compensate for the otherwise excessive annihilation rates. As a result, regions of the IDM parameter space previously deemed non-viable become compatible with the observed DM abundance without requiring additional interactions or modifications of the particle content.

Our findings highlight that conclusions drawn about the viability of particle DM models cannot be disentangled from assumptions concerning the pre–Big Bang nucleosynthesis expansion history of the Universe. More broadly, they underscore the importance of extending DM phenomenology beyond the standard cosmological paradigm, particularly in well-motivated theories where deviations such as stiff eras naturally arise. The IDM thus provides a clear and minimal example of how non-standard cosmological histories can reopen and reshape the DM parameter space, motivating further studies at the interface of particle physics and early-Universe cosmology.

A similar analysis can be straightforwardly applied to other well-motivated WIMP candidates -- such as wino-like DM -- the relic abundance of which is also strongly constrained by efficient gauge-mediated annihilations in the standard cosmological scenario. Looking ahead, future probes of the early Universe -- including precision measurements of the CMB, searches for stochastic gravitational-wave backgrounds sensitive to non-standard expansion histories, and next-generation DM detection experiments -- will play a key role in testing such scenarios and further constraining the interplay between particle physics and cosmology.

\acknowledgments
NB received funding from the grants PID2023-151418NB-I00 funded by MCIU/AEI/10.13039 /501100011033/ FEDER and PID2022-139841NB-I00 of MICIU/AEI/10.13039/501100011033 and FEDER, UE. GB and AP  thank the  Institute For Interdisciplinary Research in Science and Education (IFIRSE), Quy Nhon, Vietnam, and {\it Rencontres du Vietnam} for their warm hospitality. The work of AP  was conducted under the state assignment of Lomonosov State University. 

\appendix

\section{Implementation in micrOMEGAs} \label{app:micromegas}
New functions are available in micrOMEGAs 7.0~\cite{micro7} to compute the DM relic density after solving the background dynamics. We keep track of the evolution of the cosmological background with the function\\
\noindent \verb|getInflDecayExt(Hsm, HI, Gamma, alpha, w, &Trh, &Tmax, &aEnd)|\\
which solves the set of Boltzmann transport equations
\begin{align}
    &\frac{d\rp}{dt} + 3\, (1 + w)\, H\, \rp = 0\,, \label{eq:rho}\\
    &\frac{ds}{dt} + 3\, H\, s = 0\,, \label{eq:s}
\end{align}
for the inflaton energy density and the SM entropy density
\begin{equation}
   s = \frac{2\pi^2}{45}\, h_\text{eff}(T)\, T^3,
\end{equation}
where $h_\text{eff}(T)$ is the number of relativistic degrees of freedom that contribute to the SM entropy. The input parameters are defined at $a = a_I = 1$: {\tt Hsm} - the  contribution  of the SM bath to the Hubble rate ($H^2_\text{SM}=\rho_R/(3M_P^2)$); \verb|HI| - the contribution of the inflaton to the  Hubble rate ($H^2_\phi=\rho_\phi/(3M_P^2)$); \verb|Gamma| - the decay width of the inflaton (here \verb|Gamma|~$ = 0$); \verb|alpha| - the scaling of the SM temperature as a function of the scale factor (irrelevant here as \verb|Gamma|~$ = 0$); and \verb|w| - the equation-of-state parameter of $\phi$ (with \verb|w|~$> 1/3$). \verb|Trh|, \verb|Tmax| and \verb|aEnd| are return parameters. They represent respectively the temperature at which the energy density of the inflaton becomes equal to the energy density of SM particles, the maximal temperature reached, and the scale factor at the temperature \verb|Tend|. This temperature is defined by the user as a global parameter. The  evolution of the Universe is stored in the  tabulated  functions Ta(a), Ha(a), rhoSMa(a), rhoIa(a) which represent respectively the temperature of the SM bath, the total expansion rate, energy  density of SM particles and energy density of the inflaton. These functions are then used to calculate the relic density of DM.  The argument of these functions belongs to the  interval $a \in [1,\, \verb|aEnd|]$. The  final Hubble scale {\tt aEnd} is defined  automatically by  the final temperature of the SM bath \verb|Tend|. The function \verb|getInflDecayExt| returns 0 when it can find a solution to the differential equations or 128 otherwise.

Even if $\Tmax$ and $\Trh$ are computed numerically, here we give analytical estimations as a function of $H_\text{SM}$ and $H_\phi$ at $a=a_I$~\cite{Barman:2021ugy, Bernal:2024yhu}
\begin{align}
    \Tmax &= \left[\frac{3}{\pi}\, \sqrt{\frac{10}{\gs}}\, M_P\, H_\text{SM}(a_I)\right]^{1/2},\\
    \Trh &\simeq \Tmax \left[\frac{H_\text{SM}(a_I)}{H_\phi(a_I)}\right]^\frac{2}{3w-1}.
\end{align}
Once the cosmological background has been computed, the function\\
\noindent \verb|darkOmegaInfl(b, MI, Beps, &Tfo, &err)|\\
solves the DM evolution Eq.~\eqref{eq:BEDM}, where {\tt b} is the net number of DM particles per decay of the  inflaton and {\tt MI} is the mass of the inflaton in GeV. Here \verb|b| and \verb|MI| are irrelevant since we assume \verb|Gamma|~$ = 0$. This function  uses Ta, rhoIa, and Ha obtained by \verb|getInflDecayExt|.  The main return value of this routine is the DM relic density $\Omega_\chi h^2$. The auxiliary output parameters are {\tt Tfo}, and  {\tt err}. The parameter {\tt Tfo} returns the freeze-out temperature, recall that in freeze-out the final relic density decreases when the annihilation cross section  increases, otherwise \verb|Tfo|~=~{\tt NAN}.\footnote{ The C-function {\it isfinite} checks the number and returns 0 in case of {\tt NAN}} The error code {\tt err} indicates whether there is a problem with the solution of the differential Eqs.~\eqref{eq:rho} and~\eqref{eq:s}, in which case the 8$^\text{th}$ bit of the error code {\tt err} is 1, or with the solution of Eq.~\eqref{eq:BEDM}, then the 9$^\text{th}$ bit is 1. The lowest bits contain information about problems with numerical integration and should be treated as warnings. The parameter \verb|Beps| has the same meaning as in other \verb|darkOmega| routines; it determines the condition to include Boltzmann suppressed channels~\cite{Belanger:2001fz}. The DM abundance is stored in an array and is available through the functions {\tt Za(a)}, \verb|Ya(a)| while the equilibrium abundance is stored in {\tt ZaEq(a)}, \verb|YaEq(a)|. 

\section{The case \boldmath \texorpdfstring{$\lambda_L \ne 0$}{}} \label{app:more}
\begin{figure}[t!]
    \def\sepf{0.496}
    \centering
    \includegraphics[width=\sepf\columnwidth]{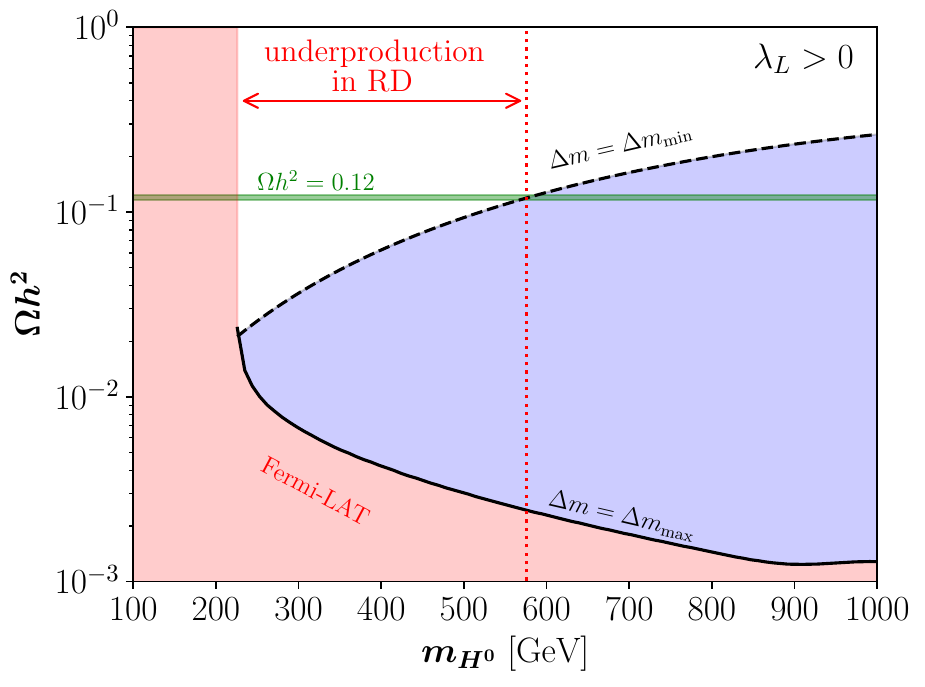}
    \includegraphics[width=\sepf\columnwidth]{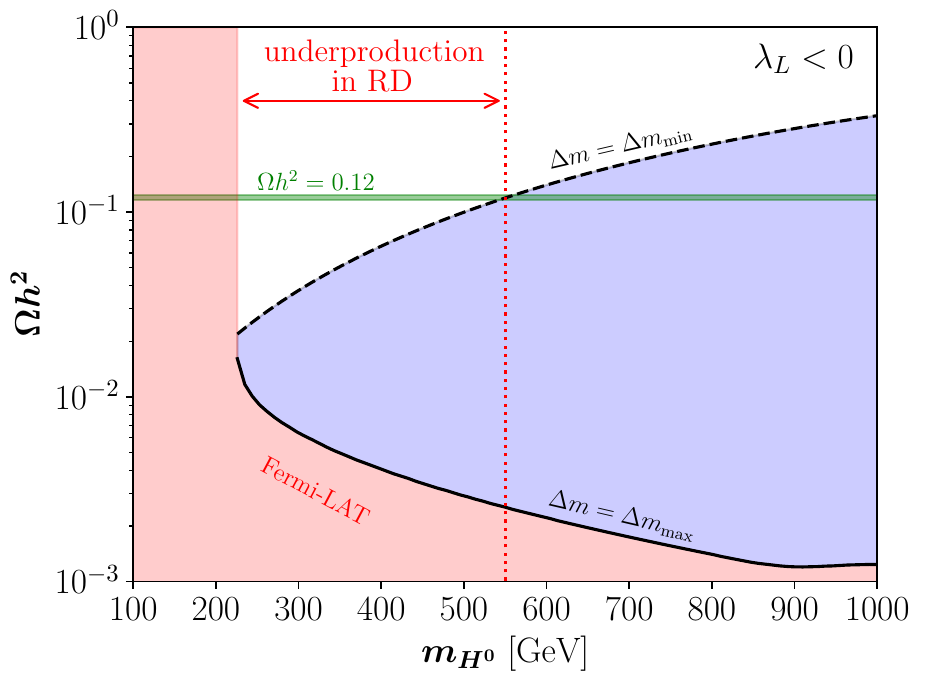}
    \includegraphics[width=\sepf\columnwidth]{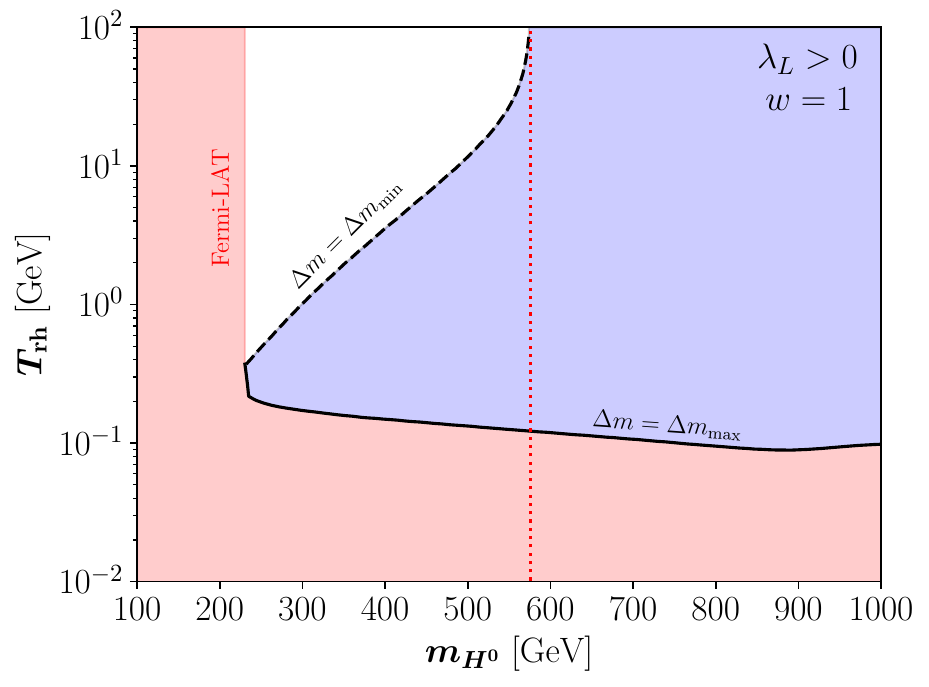}
    \includegraphics[width=\sepf\columnwidth]{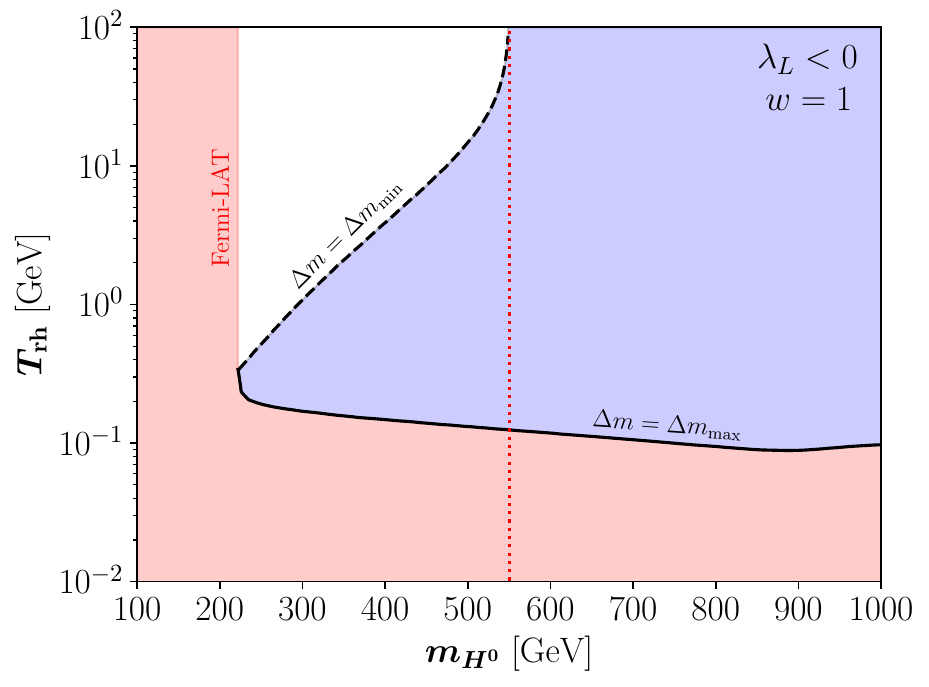}
    \caption{Same as Figs.~\ref{fig:RD} and~\ref{fig:kination} but for $\lambda_L > 0$ (left) or $\lambda_L < 0$ (right). Note that in the latter case, the requirement of vacuum stability essentially excludes the blue-shaded region for masses $m_{H^0} \gtrsim 600$ GeV.}
    \label{fig:kination2}
\end{figure}

In Section~\ref{sec:DM} we focused on the case of $\lambda_L = 0$. For completeness, here we present the results corresponding to the case $\lambda_L \ne 0$, even if they are found to not introduce qualitatively new features. In Fig.~\ref{fig:kination2} we show the corresponding results of Figs.~\ref{fig:RD} and ~\ref{fig:kination}, for $\lambda_L > 0$ (left) and $\lambda_L < 0$ (right). We recall that in both cases the absolute value of $\lambda_L$ is fixed to its maximum value allowed by direct detection. As  mentioned in Section~\ref{sec:idm}, since the current DD limits force $\lambda_L$ to be small, in these cases too the processes that control  the relic density  are annihilations into gauge bosons (along with coannihilation when $\Delta m$ is small), whereas annihilations into the final states $hh$ and $t\bar t$ only provide a subdominant contribution. 

Due to the constructive interference between the Higgs exchange diagram in $H^0H^0\to W^+W^-(ZZ)$ and the 4-point vertex ($H^0H^0W^+W^-$) for $\lambda_L >0$~\cite{LopezHonorez:2010tb}, the annihilation cross-section increases mildly and thus the relic density decreases, and conversely for $\lambda_L <0$, as can be seen in the upper panels of Fig.~\ref{fig:kination2}. The  impact of varying $\lambda_L$ is almost imperceptible  when $\Delta m$ is small because co-annihilation channels, which provide a substantial contribution to the relic density, are mostly independent of $\lambda_L$.

Lastly, we should note that in the case of $\lambda_L < 0$, the vacuum stability constraint has an impact on part of the parameter space depicted in Fig.~\ref{fig:kination2} (right). Concretely, we find that for these scenarios the electroweak vacuum becomes unstable for DM masses roughly above $600$~GeV for the entire $\Delta m$ range that we consider. However, the mass range between $\sim 200$~GeV and $\sim 550$~GeV which becomes viable when a kination period is introduced remains immune to these constraints.

\section{Maximum temperature of the SM thermal bath} \label{app:Tmax}
\begin{figure}[t!]
    \def\sepf{0.496}
    \centering
    \includegraphics[width=\sepf\columnwidth]{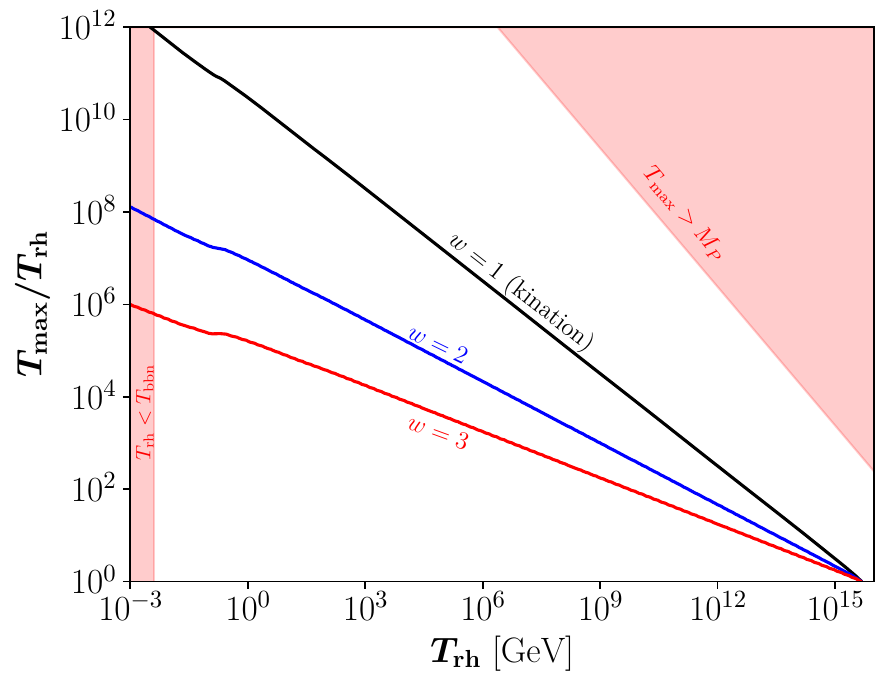}
    \caption{Maximum value of $\Tmax$ allowed by the inflationary scale $H_I$ as a function of $\Trh$, for different values of $w$.}
    \label{fig:TmaxTrh}
\end{figure} 
A limit on the maximum temperature $\Tmax$ reached by the SM thermal bath can be derived by taking into account the upper bound on the inflationary scale $H_I \simeq 2.0 \times 10^{-5}~M_P$~\cite{BICEP:2021xfz}. Comparing the Hubble expansion rates at the end of inflation (that is, at $T = \Tmax$), one obtains
\begin{align}
    3\, M_P^2\, H_I^2 &\gtrsim \rp(\Tmax) + \rR(\Tmax) \nonumber\\
    &\gtrsim \frac{\pi^2}{30}\, \gs(\Trh)\, \Trh^4 \left[\left(\frac{\gss(\Tmax)}{\gss(\Trh)}\right)^{\omega+1} \left(\frac{\Tmax}{\Trh}\right)^{3(\omega+1)} + \frac{\gs(\Tmax)}{\gs(\Trh)} \left(\frac{\Tmax}{\Trh}\right)^4\right].
\end{align}
The corresponding upper bounds for $\omega = 1$, 2, and 3 are shown in Fig.~\ref{fig:TmaxTrh}.

\bibliographystyle{JHEP}
\bibliography{biblio}

@article{Ma:2006km,
    author = "Ma, Ernest",
    title = "{Verifiable radiative seesaw mechanism of neutrino mass and dark matter}",
    eprint = "hep-ph/0601225",
    archivePrefix = "arXiv",
    reportNumber = "UCRHEP-T403",
    doi = "10.1103/PhysRevD.73.077301",
    journal = "Phys. Rev. D",
    volume = "73",
    pages = "077301",
    year = "2006"
}

@article{Barbieri:2006dq,
    author = "Barbieri, Riccardo and Hall, Lawrence J. and Rychkov, Vyacheslav S.",
    title = "{Improved naturalness with a heavy Higgs: An Alternative road to LHC physics}",
    eprint = "hep-ph/0603188",
    archivePrefix = "arXiv",
    reportNumber = "UCB-PTH-06-04, LBNL-59894",
    doi = "10.1103/PhysRevD.74.015007",
    journal = "Phys. Rev. D",
    volume = "74",
    pages = "015007",
    year = "2006"
}

@article{Deshpande:1977rw,
    author = "Deshpande, Nilendra G. and Ma, Ernest",
    title = "{Pattern of Symmetry Breaking with Two Higgs Doublets}",
    reportNumber = "OITS-81",
    doi = "10.1103/PhysRevD.18.2574",
    journal = "Phys. Rev. D",
    volume = "18",
    pages = "2574",
    year = "1978"
}

@article{LopezHonorez:2006gr,
    author = "L\'opez Honorez, Laura and Nezri, Emmanuel and Oliver, Josep F. and Tytgat, Michel H. G.",
    title = "{The Inert Doublet Model: An Archetype for Dark Matter}",
    eprint = "hep-ph/0612275",
    archivePrefix = "arXiv",
    reportNumber = "ULB-TH-06-27",
    doi = "10.1088/1475-7516/2007/02/028",
    journal = "JCAP",
    volume = "02",
    pages = "028",
    year = "2007"
}

@article{LopezHonorez:2010tb,
    author = "L\'opez Honorez, Laura and Yaguna, Carlos E.",
    title = "{A new viable region of the inert doublet model}",
    eprint = "1011.1411",
    archivePrefix = "arXiv",
    primaryClass = "hep-ph",
    reportNumber = "ULB-TH-10-37",
    doi = "10.1088/1475-7516/2011/01/002",
    journal = "JCAP",
    volume = "01",
    pages = "002",
    year = "2011"
}

@article{Goudelis:2013uca,
    author = "Goudelis, A. and Herrmann, B. and St{\r{a}}l, O.",
    title = "{Dark matter in the Inert Doublet Model after the discovery of a Higgs-like boson at the LHC}",
    eprint = "1303.3010",
    archivePrefix = "arXiv",
    primaryClass = "hep-ph",
    reportNumber = "LAPTH-006-13",
    doi = "10.1007/JHEP09(2013)106",
    journal = "JHEP",
    volume = "09",
    pages = "106",
    year = "2013"
}

@article{Eiteneuer:2017hoh,
    author = "Eiteneuer, Benedikt and Goudelis, Andreas and Heisig, Jan",
    title = "{The inert doublet model in the light of Fermi-LAT gamma-ray data: a global fit analysis}",
    eprint = "1705.01458",
    archivePrefix = "arXiv",
    primaryClass = "hep-ph",
    reportNumber = "TTK-17-12",
    doi = "10.1140/epjc/s10052-017-5166-1",
    journal = "Eur. Phys. J. C",
    volume = "77",
    number = "9",
    pages = "624",
    year = "2017"
}

@article{Lundstrom:2008ai,
    author = "Lundstr{\"o}m, Erik and Gustafsson, Michael and Edsj{\"o}, Joakim",
    title = "{The Inert Doublet Model and LEP II Limits}",
    eprint = "0810.3924",
    archivePrefix = "arXiv",
    primaryClass = "hep-ph",
    doi = "10.1103/PhysRevD.79.035013",
    journal = "Phys. Rev. D",
    volume = "79",
    pages = "035013",
    year = "2009"
}

@article{Pierce:2007ut,
    author = "Pierce, Aaron and Thaler, Jesse",
    title = "{Natural Dark Matter from an Unnatural Higgs Boson and New Colored Particles at the TeV Scale}",
    eprint = "hep-ph/0703056",
    archivePrefix = "arXiv",
    reportNumber = "MCTP-07-11",
    doi = "10.1088/1126-6708/2007/08/026",
    journal = "JHEP",
    volume = "08",
    pages = "026",
    year = "2007"
}

@article{Eriksson:2009ws,
    author = "Eriksson, David and Rathsman, Johan and Stål, Oscar",
    title = "{2HDMC: Two-Higgs-Doublet Model Calculator Physics and Manual}",
    eprint = "0902.0851",
    archivePrefix = "arXiv",
    primaryClass = "hep-ph",
    doi = "10.1016/j.cpc.2009.09.011",
    journal = "Comput. Phys. Commun.",
    volume = "181",
    pages = "189--205",
    year = "2010"
}

@article{Planck:2018vyg,
    author = "Aghanim, N. and others",
    collaboration = "Planck",
    title = "{Planck 2018 results. VI. Cosmological parameters}",
    eprint = "1807.06209",
    archivePrefix = "arXiv",
    primaryClass = "astro-ph.CO",
    doi = "10.1051/0004-6361/201833910",
    journal = "Astron. Astrophys.",
    volume = "641",
    pages = "A6",
    year = "2020",
    note = "[Erratum: Astron.Astrophys. 652, C4 (2021)]"
}

@article{BICEP:2021xfz,
    author = "Ade, P. A. R. and others",
    collaboration = "BICEP, Keck",
    title = "{Improved Constraints on Primordial Gravitational Waves using Planck, WMAP, and BICEP/Keck Observations through the 2018 Observing Season}",
    eprint = "2110.00483",
    archivePrefix = "arXiv",
    primaryClass = "astro-ph.CO",
    doi = "10.1103/PhysRevLett.127.151301",
    journal = "Phys. Rev. Lett.",
    volume = "127",
    number = "15",
    pages = "151301",
    year = "2021"
}

@article{Kofman:1997yn,
    author = "Kofman, Lev and Linde, Andrei D. and Starobinsky, Alexei A.",
    title = "{Towards the theory of reheating after inflation}",
    eprint = "hep-ph/9704452",
    archivePrefix = "arXiv",
    reportNumber = "IFA-97-28, SU-ITP-97-18",
    doi = "10.1103/PhysRevD.56.3258",
    journal = "Phys. Rev. D",
    volume = "56",
    pages = "3258--3295",
    year = "1997"
}

@article{Hannestad:2004px,
    author = "Hannestad, Steen",
    title = "{What is the lowest possible reheating temperature?}",
    eprint = "astro-ph/0403291",
    archivePrefix = "arXiv",
    doi = "10.1103/PhysRevD.70.043506",
    journal = "Phys. Rev. D",
    volume = "70",
    pages = "043506",
    year = "2004"
}

@article{Kawasaki:2000en,
    author = "Kawasaki, M. and Kohri, Kazunori and Sugiyama, Naoshi",
    title = "{MeV scale reheating temperature and thermalization of neutrino background}",
    eprint = "astro-ph/0002127",
    archivePrefix = "arXiv",
    doi = "10.1103/PhysRevD.62.023506",
    journal = "Phys. Rev. D",
    volume = "62",
    pages = "023506",
    year = "2000"
}

@article{deSalas:2015glj,
    author = "de Salas, P. F. and Lattanzi, M. and Mangano, G. and Miele, G. and Pastor, S. and Pisanti, O.",
    title = "{Bounds on very low reheating scenarios after Planck}",
    eprint = "1511.00672",
    archivePrefix = "arXiv",
    primaryClass = "astro-ph.CO",
    reportNumber = "IFIC-15-70",
    doi = "10.1103/PhysRevD.92.123534",
    journal = "Phys. Rev. D",
    volume = "92",
    number = "12",
    pages = "123534",
    year = "2015"
}

@article{Barman:2021ugy,
    author = "Barman, Basabendu and Bernal, Nicol\'as",
    title = "{Gravitational SIMPs}",
    eprint = "2104.10699",
    archivePrefix = "arXiv",
    primaryClass = "hep-ph",
    reportNumber = "PI/UAN-2021-688FT",
    doi = "10.1088/1475-7516/2021/06/011",
    journal = "JCAP",
    volume = "06",
    pages = "011",
    year = "2021"
}

@article{Alguero:2023zol,
    author = "Alguero, G. and B\'elanger, G. and Boudjema, F. and Chakraborti, S. and Goudelis, A. and Kraml, S. and Mjallal, A. and Pukhov, A.",
    title = "{micrOMEGAs 6.0: N-component dark matter}",
    eprint = "2312.14894",
    archivePrefix = "arXiv",
    primaryClass = "hep-ph",
    doi = "10.1016/j.cpc.2024.109133",
    journal = "Comput. Phys. Commun.",
    volume = "299",
    pages = "109133",
    year = "2024"
}

@unpublished{micro7,
    author = "B\'elanger, G. and Belyaev, Alexander and Bernal, Nicolas and Boudjema, F. and Chakraborti, S. and Goudelis, A. and Pukhov, A.",
    title = "{micrOMEGAs 7.0: Non-standard cosmology}",
    year = "To appear"
}

@article{Bernal:2024yhu,
    author = "Bernal, Nicol\'as and Deka, Kuldeep and Losada, Marta",
    title = "{Thermal dark matter with low-temperature reheating}",
    eprint = "2406.17039",
    archivePrefix = "arXiv",
    primaryClass = "hep-ph",
    doi = "10.1088/1475-7516/2024/09/024",
    journal = "JCAP",
    volume = "09",
    pages = "024",
    year = "2024"
}

@article{Batell:2024dsi,
    author = "Batell, Brian and others",
    title = "{Conversations and deliberations: Non-standard cosmological epochs and expansion histories}",
    eprint = "2411.04780",
    archivePrefix = "arXiv",
    primaryClass = "astro-ph.CO",
    doi = "10.1142/S0217751X25300042",
    journal = "Int. J. Mod. Phys. A",
    volume = "40",
    number = "17",
    pages = "2530004",
    year = "2025"
}

@article{Allahverdi:2020bys,
    author = "Allahverdi, Rouzbeh and others",
    title = "{The First Three Seconds: a Review of Possible Expansion Histories of the Early Universe}",
    eprint = "2006.16182",
    archivePrefix = "arXiv",
    primaryClass = "astro-ph.CO",
    reportNumber = "FERMILAB-PUB-20-242-A, KCL-PH-TH/2020-33, KEK-Cosmo-257,
  KEK-TH-2231, IPMU20-0070, PI/UAN-2020-674FT, RUP-20-22",
    doi = "10.21105/astro.2006.16182",
    journal = "Open J.Astrophys.",
    volume = "4",
    month = "6",
    year = "2021"
}

@article{Belanger:2001fz,
    author = "B\'elanger, G. and Boudjema, F. and Pukhov, A. and Semenov, A.",
    title = "{MicrOMEGAs: A Program for calculating the relic density in the MSSM}",
    eprint = "hep-ph/0112278",
    archivePrefix = "arXiv",
    reportNumber = "LAPTH-881-01",
    doi = "10.1016/S0010-4655(02)00596-9",
    journal = "Comput. Phys. Commun.",
    volume = "149",
    pages = "103--120",
    year = "2002"
}

@article{Spokoiny:1993kt,
    author = "Spokoiny, Boris",
    title = "{Deflationary universe scenario}",
    eprint = "gr-qc/9306008",
    archivePrefix = "arXiv",
    reportNumber = "KUNS-1201",
    doi = "10.1016/0370-2693(93)90155-B",
    journal = "Phys. Lett. B",
    volume = "315",
    pages = "40--45",
    year = "1993"
}

@article{Ferreira:1997hj,
    author = "Ferreira, Pedro G. and Joyce, Michael",
    title = "{Cosmology with a primordial scaling field}",
    eprint = "astro-ph/9711102",
    archivePrefix = "arXiv",
    reportNumber = "CFPA-97-TH-20",
    doi = "10.1103/PhysRevD.58.023503",
    journal = "Phys. Rev. D",
    volume = "58",
    pages = "023503",
    year = "1998"
}

@article{Heisig:2018kfq,
    author = "Heisig, Jan and Kraml, Sabine and Lessa, Andre",
    title = "{Constraining new physics with searches for long-lived particles: Implementation into SModelS}",
    eprint = "1808.05229",
    archivePrefix = "arXiv",
    primaryClass = "hep-ph",
    reportNumber = "TTK-18-31",
    doi = "10.1016/j.physletb.2018.10.049",
    journal = "Phys. Lett. B",
    volume = "788",
    pages = "87--95",
    year = "2019"
}

@article{Khoury:2001wf,
    author = "Khoury, Justin and Ovrut, Burt A. and Steinhardt, Paul J. and Turok, Neil",
    title = "{The Ekpyrotic universe: Colliding branes and the origin of the hot big bang}",
    eprint = "hep-th/0103239",
    archivePrefix = "arXiv",
    doi = "10.1103/PhysRevD.64.123522",
    journal = "Phys. Rev. D",
    volume = "64",
    pages = "123522",
    year = "2001"
}

@article{Choi:1999xn,
    author = "Choi, Kiwoon",
    title = "{String or M theory axion as a quintessence}",
    eprint = "hep-ph/9902292",
    archivePrefix = "arXiv",
    reportNumber = "KAIST-TH-99-03",
    doi = "10.1103/PhysRevD.62.043509",
    journal = "Phys. Rev. D",
    volume = "62",
    pages = "043509",
    year = "2000"
}

@article{Alguero:2021dig,
    author = {Alguero, Ga{\"e}l and Heisig, Jan and Khosa, Charanjit K. and Kraml, Sabine and Kulkarni, Suchita and Lessa, Andre and Reyes-Gonz{\'a}lez, Humberto and Waltenberger, Wolfgang and Wongel, Alicia},
    title = "{Constraining new physics with SModelS version 2}",
    eprint = "2112.00769",
    archivePrefix = "arXiv",
    primaryClass = "hep-ph",
    reportNumber = "TTK-21-50",
    doi = "10.1007/JHEP08(2022)068",
    journal = "JHEP",
    volume = "08",
    pages = "068",
    year = "2022"
}

@article{Gardner:2004in,
    author = "Gardner, Carl L.",
    title = "{Quintessence and the transition to an accelerating universe}",
    eprint = "astro-ph/0407604",
    archivePrefix = "arXiv",
    doi = "10.1016/j.nuclphysb.2004.11.065",
    journal = "Nucl. Phys. B",
    volume = "707",
    pages = "278--300",
    year = "2005"
}

@article{DEramo:2017gpl,
    author = "D'Eramo, Francesco and Fern\'andez, Nicolas and Profumo, Stefano",
    title = "{When the Universe Expands Too Fast: Relentless Dark Matter}",
    eprint = "1703.04793",
    archivePrefix = "arXiv",
    primaryClass = "hep-ph",
    reportNumber = "SCIPP-17-02",
    doi = "10.1088/1475-7516/2017/05/012",
    journal = "JCAP",
    volume = "05",
    pages = "012",
    year = "2017"
}

@article{Cyburt:2015mya,
    author = "Cyburt, Richard H. and Fields, Brian D. and Olive, Keith A. and Yeh, Tsung-Han",
    title = "{Big Bang Nucleosynthesis: 2015}",
    eprint = "1505.01076",
    archivePrefix = "arXiv",
    primaryClass = "astro-ph.CO",
    reportNumber = "UMN-TH-3432-15, FTPI-MINN-15-19",
    doi = "10.1103/RevModPhys.88.015004",
    journal = "Rev. Mod. Phys.",
    volume = "88",
    pages = "015004",
    year = "2016"
}

@article{LZ:2024zvo,
    author = "Aalbers, J. and others",
    collaboration = "LZ",
    title = "{Dark Matter Search Results from 4.2{\,}{\,}Tonne-Years of Exposure of the LUX-ZEPLIN (LZ) Experiment}",
    eprint = "2410.17036",
    archivePrefix = "arXiv",
    primaryClass = "hep-ex",
    reportNumber = "FERMILAB-PUB-24-0796-V",
    doi = "10.1103/4dyc-z8zf",
    journal = "Phys. Rev. Lett.",
    volume = "135",
    number = "1",
    pages = "011802",
    year = "2025"
}

@article{Alvarez:2020cmw,
    author = "\'Alvarez, Alexandre and Calore, Francesca and Genina, Anna and Read, Justin and Serpico, Pasquale Dario and Zald\'ivar, Bryan",
    title = "{Dark matter constraints from dwarf galaxies with data-driven J-factors}",
    eprint = "2002.01229",
    archivePrefix = "arXiv",
    primaryClass = "astro-ph.HE",
    reportNumber = "LAPTH-002/20, IFT-UAM/CSIC-20-15",
    doi = "10.1088/1475-7516/2020/09/004",
    journal = "JCAP",
    volume = "09",
    pages = "004",
    year = "2020"
}

@article{Calore:2018sdx,
    author = "Calore, Francesca and Serpico, Pasquale D. and Zald\'ivar, Bryan",
    title = "{Dark matter constraints from dwarf galaxies: a data-driven analysis}",
    eprint = "1803.05508",
    archivePrefix = "arXiv",
    primaryClass = "astro-ph.HE",
    doi = "10.1088/1475-7516/2018/10/029",
    journal = "JCAP",
    volume = "10",
    pages = "029",
    year = "2018"
}

@software{calore_2021_5592836,
  author       = {Calore, Francesca and
                  Zald\'ivar, Bryan and
                  Serpico, Pasquale and
                  Eckner, Christopher},
  title        = {Dark matter constraints from dwarf galaxies: a
                   data-driven LAT analysis
                  },
  month        = oct,
  year         = 2021,
  publisher    = {Zenodo},
  version      = {v1.0},
  doi          = {10.5281/zenodo.5592836},
  url          = {https://doi.org/10.5281/zenodo.5592836},
}

@article{Navarro:1996gj,
    author = "Navarro, Julio F. and Frenk, Carlos S. and White, Simon D. M.",
    title = "{A Universal density profile from hierarchical clustering}",
    eprint = "astro-ph/9611107",
    archivePrefix = "arXiv",
    doi = "10.1086/304888",
    journal = "Astrophys. J.",
    volume = "490",
    pages = "493--508",
    year = "1997"
}

@article{Fermi-LAT:2013sme,
    author = "Ackermann, M. and others",
    collaboration = "Fermi-LAT",
    title = "{Dark Matter Constraints from Observations of 25 Milky Way Satellite Galaxies with the Fermi Large Area Telescope}",
    eprint = "1310.0828",
    archivePrefix = "arXiv",
    primaryClass = "astro-ph.HE",
    reportNumber = "FERMILAB-PUB-13-683-A",
    doi = "10.1103/PhysRevD.89.042001",
    journal = "Phys. Rev. D",
    volume = "89",
    pages = "042001",
    year = "2014"
}

@article{Zeldovich:1961sbr,
    author = "Zel'dovich, Ya. B.",
    title = "{The Equation of State at Ultrahigh Densities and Its Relativistic Limitations}",
    journal = "Zh. Eksp. Teor. Fiz.",
    volume = "41",
    pages = "1609--1615",
    year = "1961"
}

@article{Joyce:1996cp,
    author = "Joyce, Michael",
    title = "{Electroweak Baryogenesis and the Expansion Rate of the Universe}",
    eprint = "hep-ph/9606223",
    archivePrefix = "arXiv",
    reportNumber = "CERN-TH-96-098, CERN-TH-96-98",
    doi = "10.1103/PhysRevD.55.1875",
    journal = "Phys. Rev. D",
    volume = "55",
    pages = "1875--1878",
    year = "1997"
}

@article{Arcadi:2024ukq,
    author = "Arcadi, Giorgio and Cabo-Almeida, David and Dutra, Ma{\'\i}ra and Ghosh, Pradipta and Lindner, Manfred and Mambrini, Yann and Neto, Jacinto P. and Pierre, Mathias and Profumo, Stefano and Queiroz, Farinaldo S.",
    title = "{The Waning of the WIMP: Endgame?}",
    eprint = "2403.15860",
    archivePrefix = "arXiv",
    primaryClass = "hep-ph",
    doi = "10.1140/epjc/s10052-024-13672-y",
    journal = "Eur. Phys. J. C",
    volume = "85",
    number = "2",
    pages = "152",
    year = "2025"
}

@article{Co:2019jts,
    author = "Co, Raymond T. and Hall, Lawrence J. and Harigaya, Keisuke",
    title = "{Axion Kinetic Misalignment Mechanism}",
    eprint = "1910.14152",
    archivePrefix = "arXiv",
    primaryClass = "hep-ph",
    reportNumber = "LCTP-19-28",
    doi = "10.1103/PhysRevLett.124.251802",
    journal = "Phys. Rev. Lett.",
    volume = "124",
    number = "25",
    pages = "251802",
    year = "2020"
}

@article{Chang:2019tvx,
    author = "Chang, Chia-Feng and Cui, Yanou",
    title = "{New Perspectives on Axion Misalignment Mechanism}",
    eprint = "1911.11885",
    archivePrefix = "arXiv",
    primaryClass = "hep-ph",
    doi = "10.1103/PhysRevD.102.015003",
    journal = "Phys. Rev. D",
    volume = "102",
    number = "1",
    pages = "015003",
    year = "2020"
}

@article{Barman:2021rdr,
    author = "Barman, Basabendu and Bernal, Nicol{\'a}s and Ramberg, Nicklas and Visinelli, Luca",
    title = "{QCD Axion Kinetic Misalignment without Prejudice}",
    eprint = "2111.03677",
    archivePrefix = "arXiv",
    primaryClass = "hep-ph",
    reportNumber = "PI/UAN-2021-703FT, MITP-21-057",
    doi = "10.3390/universe8120634",
    journal = "Universe",
    volume = "8",
    number = "12",
    pages = "634",
    year = "2022"
}

@article{Dolgov:1989us,
    author = "Dolgov, A. D. and Kirilova, D. P.",
    title = "{On Particle Creation by a Time Dependent Scalar Field}",
    reportNumber = "JINR-E2-89-321",
    journal = "Sov. J. Nucl. Phys.",
    volume = "51",
    pages = "172--177",
    year = "1990"
}

@article{Traschen:1990sw,
    author = "Traschen, Jennie H. and Brandenberger, Robert H.",
    title = "{Particle Production During Out-of-equilibrium Phase Transitions}",
    reportNumber = "BROWN-HET-731",
    doi = "10.1103/PhysRevD.42.2491",
    journal = "Phys. Rev. D",
    volume = "42",
    pages = "2491--2504",
    year = "1990"
}

@article{Kofman:1994rk,
    author = "Kofman, Lev and Linde, Andrei D. and Starobinsky, Alexei A.",
    title = "{Reheating after inflation}",
    eprint = "hep-th/9405187",
    archivePrefix = "arXiv",
    reportNumber = "UH-IFA-94-35, SU-ITP-94-13, YITP-U-94-15",
    doi = "10.1103/PhysRevLett.73.3195",
    journal = "Phys. Rev. Lett.",
    volume = "73",
    pages = "3195--3198",
    year = "1994"
}

@article{Barman:2025lvk,
    author = "Barman, Basabendu and Bernal, Nicol{\'a}s and Rubio, Javier",
    title = "{Two or three things particle physicists (mis)understand about (pre)heating}",
    eprint = "2503.19980",
    archivePrefix = "arXiv",
    primaryClass = "hep-ph",
    reportNumber = "IPARCOS-UCM-25-019",
    doi = "10.1016/j.nuclphysb.2025.116996",
    journal = "Nucl. Phys. B",
    volume = "1018",
    pages = "116996",
    year = "2025"
}

@article{Peebles:1998qn,
    author = "Peebles, P. J. E. and Vilenkin, A.",
    title = "{Quintessential inflation}",
    eprint = "astro-ph/9810509",
    archivePrefix = "arXiv",
    doi = "10.1103/PhysRevD.59.063505",
    journal = "Phys. Rev. D",
    volume = "59",
    pages = "063505",
    year = "1999"
}

@article{Bernal:2020bfj,
    author = {Bernal, Nicol{\'a}s and Rubio, Javier and Veerm{\"a}e, Hardi},
    title = "{Boosting Ultraviolet Freeze-in in NO Models}",
    eprint = "2004.13706",
    archivePrefix = "arXiv",
    primaryClass = "hep-ph",
    reportNumber = "PI/UAN-2020-670FT, HIP-2020-12/TH",
    doi = "10.1088/1475-7516/2020/06/047",
    journal = "JCAP",
    volume = "06",
    pages = "047",
    year = "2020"
}

@article{Klasen:2013btp,
    author = "Klasen, Michael and Yaguna, Carlos E. and Ruiz-\'Alvarez, Jose D.",
    title = "{Electroweak corrections to the direct detection cross section of inert higgs dark matter}",
    eprint = "1302.1657",
    archivePrefix = "arXiv",
    primaryClass = "hep-ph",
    reportNumber = "MS-TP-13-01",
    doi = "10.1103/PhysRevD.87.075025",
    journal = "Phys. Rev. D",
    volume = "87",
    pages = "075025",
    year = "2013"
}
\end{document}